# Class-Based Service Connectivity using Multi-Level Bandwidth Adaptation in Multimedia Wireless Networks

Mostafa Zaman Chowdhury[1] and Yeong Min Jang[2]

[1]Department of Electrical and Electronic Engineering, Khulna University of Engineering and Technology, Bangladesh

[2]Department of Electronics Engineering, Kookmin University, Korea.

E-mail: mzceee@yahoo.com, yjang@kookmin.ac.kr

**Abstract** Due to the fact that quality of service requirements are not very strict for all traffic types, more calls of higher priority can be accommodated by reducing some bandwidth allocation for the bandwidth adaptive calls. The bandwidth adaptation to accept a higher priority call is more than that of a lower priority call. Therefore, the multi-level bandwidth adaptation technique improves the overall forced call termination probability as well as provides priority of the traffic classes in terms of call blocking probability without reducing the bandwidth utilization. We propose a novel bandwidth adaptation model that releases multi-level of bandwidth from the existing multimedia traffic calls. The amount of released bandwidth is decided based on the priority of the requesting traffic calls and the number of existing bandwidth adaptive calls. This prioritization of traffic classes does not reduce the bandwidth utilization. Moreover, our scheme reduces the overall forced call termination probability significantly. The proposed scheme is modeled using the Markov Chain. The numerical results show that the proposed scheme is able to provide negligible handover call dropping probability as well as significantly reduced new call blocking probability of higher priority calls without increasing the overall forced call termination probability.

## 1. Introduction

Miscellaneous types of traffic related to a variety of multimedia applications are supported by existing wireless communication systems. A variety of multimedia applications increases the traffic load on wireless networks. The bandwidth resources of the existing wireless technologies are not sufficient to support all these traffic together during busy hours. Among the variety of traffic, few types of traffic calls e.g., traffic related to security, healthcare, and banking are more important than others [1]. Therefore, the resource management scheme should assure a satisfied service level (e.g., negligible call blocking probability) for these important classes of traffic calls without reducing the bandwidth utilization. Moreover, wireless network systems generate huge amount of handover calls. These handover calls should also be guaranteed in terms of negligible handover call dropping probability. When a higher priority call occurs, there is a possibility that, due to limited resources in the system, the call will be blocked. From a user's point of view, blocking a low priority calls is more preferable than blocking a high priority one. Therefore, of interest are mechanisms that would allow reduction in the call blocking probability of higher priority call without reducing the bandwidth utilization, even if this reduction comes at the expense of increasing the call blocking probabilities of lower priority calls. Numerous prior researches have already been done to allow higher priority for handover calls over new calls [2, 3]. The schemes based on guard bands give priority for higher priority calls by reserving few bandwidth for these calls. These schemes can reduce the call blocking probability for the higher priority calls but they also result in reduced bandwidth utilization. The flexible quality of service (QoS) provisioning has been proposed by several researchers [2, 4] to reduce the call blocking probability. The QoS adaptation techniques [5, 6] proved more flexible and efficient in guaranteeing QoS than the guard channel schemes [2]. D. D. Vergados et al. [2] proposed an adaptive resource allocation scheme to prioritize particular traffic classes over others. His proposed scheme is based on the QoS degradation of only low priority traffic to accept higher priority traffic call requests. W. Zhuang et al. [4] proposed an adaptive QoS scheme which reduces the QoS levels of calls that carry adaptive traffic to accept the handover call requests only.

The QoS provisioning is the main concern in wireless communication systems. The QoS requirements for all the traffic types are not same. Few traffic types require guaranteed bit rate (GBR), while others are categorized as "best effort" delivery only. The streaming services are limited to the delay variation of the end-to-end flow while the background services are delay insensitive. Typically, real-time services require GBR. However, non-real-time services do not necessitate GBR. Thus, under heavy traffic condition, the bandwidth of non-real-time services can be purposely degraded, so that the QoS of real-time services is preserved by maintaining low probability of blocking new calls or low probability of handover dropping calls. We present a novel bandwidth adaptation model which allows releasing of some bandwidth from the bandwidth adaptive calls, to accept higher priority calls as well as lower priority calls, when the system's resources are running low. Therefore, the scheme can reduce the overall forced call termination probability significantly as well as the call blocking probability for the higher priority calls. The features of our proposed scheme include the priority of traffic calls to provide reduced call blocking probability for higher priority traffic calls, handover priority to provide negligible handover call dropping probability, and bandwidth adaptation to increase the number of call admission in the system. The proposed scheme reserves some releasable bandwidth to accept higher priority calls by providing multi-level bandwidth adaptation. In particular, our scheme is based on *M* traffic classes and handover calls, where *M+1* bandwidth-degradation thresholds are defined for each traffic class. These thresholds signify the maximum portion of the allocated bandwidth that can be released from a bandwidth adaptive call of a particular traffic class. Our scheme allows releasing more bandwidth in the case of higher priority calls, thus increasing the probability of accepting higher priority calls, as opposed to lower priority calls.

The rest of this paper is organized as follows. Section 2 presents the system model for the proposed scheme. The Markov Chain for the proposed scheme is also presented in this section. Numerical performance evaluation results of the proposed scheme are presented and compared with other schemes in Section 3. Finally, Section 4 concludes our work.

## 2. Proposed Multi-Level Bandwidth Adaptation

The non-real-time traffic services are normally bandwidth adaptive [5-9] and, usually, do not need QoS guarantees. This traffic can tolerate bandwidth below a certain preferred bandwidth and is more adaptive to network conditions. The proposed scheme admits different types of calls in the system based on the multi-level bandwidth adaptation policy. The basic idea for the proposed bandwidth adaptation model is shown in Fig. 1. A bandwidth adaptive call of *m-th* class traffic can be allocated or re-adjusted by different level of bandwidth based on the priority of the requesting call. This figure shows the amount of maximum releasable bandwidth from each of the calls and also the minimum amount of bandwidth that should be allocated for each of the calls after accepting a call request.

The bandwidth adaptation is happened only if the empty bandwidth of the system is not enough to accept a call request. The bandwidth allocation model is characterized by bandwidth degradation factors $\gamma_m$ and $\gamma_{m,p}$, respectively, are defined for each of the *m-th* class traffic, as: the fraction of the bandwidth that has been already released from an admitted call, the maximum fraction of the bandwidth of an admitted call that can be released to accept a call of *p-th* priority traffic (*p=0* represents highest priority handover calls of any types of traffic). We classify the traffic classes in such a way that the traffic class also indicates the priority for new calls of that class (i.e., *p=m* for *m*=1 to *M)*. Therefore, to accept a handover call, the existing *m-th* class traffic call can be degraded to the maximum limit of the $\gamma_{m,0}$ portion. However, to accept a new call of *m-th* class traffic, the existing *m-th* class traffic call can be degraded to the maximum limit of the $\gamma_{m,m}$ portion. The parameter $C_{m,0}$ refers the minimum allocated bandwidth for a call of *m-th* class traffic. This reduced allocated bandwidth is permitted only to accept handover call request in the system. On the other hand, $C_{m,p}$ denotes the minimum reduced allocated bandwidth for a call of traffic class *m* to accept a new call request of class *p* in the system. Therefore, the proposed scheme gives highest priority for the handover calls (any class of traffic) and then consequently traffic class 1, class 2, …, and class *M*. Since the bandwidth of real-time traffic classes is not normally adaptive, the bandwidth degradation factor for the real-time traffic classes equals to zero. However, the system can release bandwidth from the existing non-real-time traffic calls to accept both for the non-real-time and real-time traffic calls. The level of bandwidth degradation to accept a handover call and a new call of different classes of traffic are not, necessarily, equal. Our proposed scheme can be explained with a suitable example. Suppose a system have high priority handover call, medium priority voice call, and low priority background call. According to our proposed scheme, a running background call will be accommodated with lower bandwidth to accept a handover call request compared to a new voice call request when the system bandwidth is not sufficient to accept a call request.

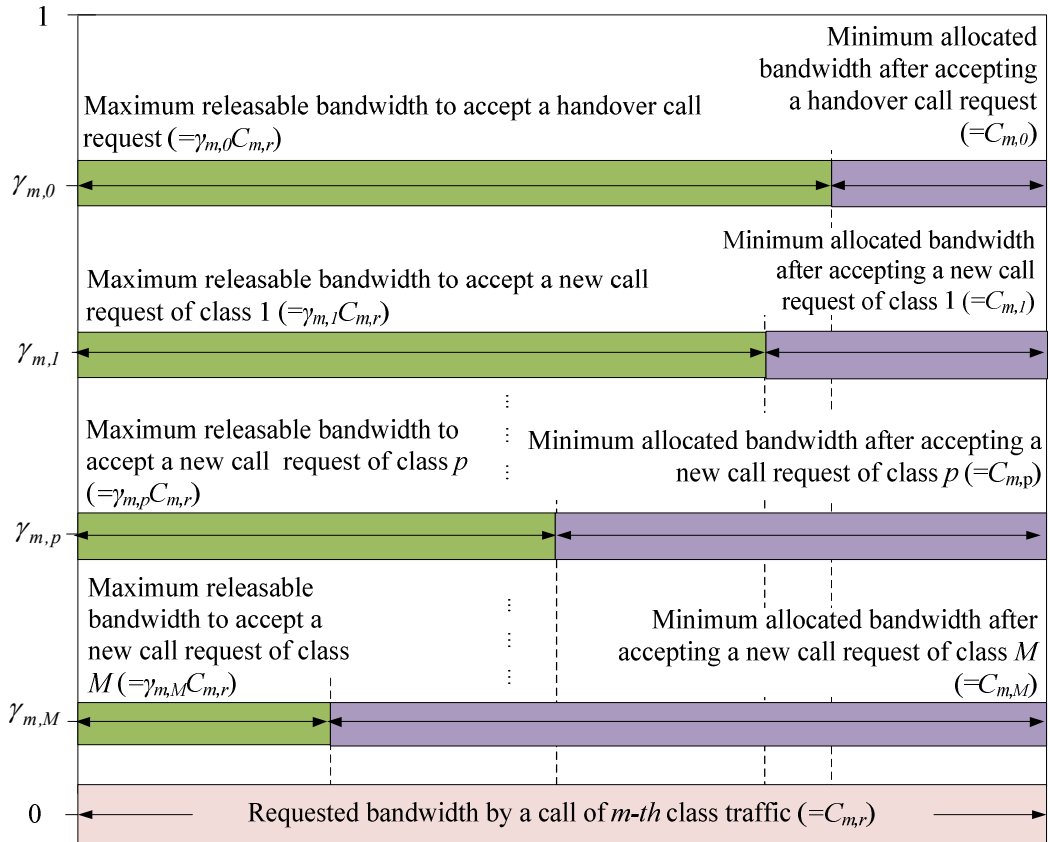

**Fig. 1** Proposed multi-level bandwidth adaptation model.

For the *m-th* class traffic calls, the required bandwidth to accept a handover call is less than that of for the new call (in case of non-adaptive QoS traffic calls it is equal). The minimum required bandwidth for a handover call is $C_{m,0}$. However, for the case of new call it is equal to $C_{m,m}$. The relation between the bandwidth-degradation factors and the bandwidth allocations are as follows

$$C_{m,a} = C_{m,r} - \gamma_m C_{m,r} \quad (1)$$

$$C_{m,p} = C_{m,r} - \gamma_{m,p} C_{m,r} \quad (2)$$

where $C_{m,r}$ and $C_{m,a}$ are, respectively, the bandwidth requested by a call and currently allocated bandwidth for each of the existing calls of the *m-th* class traffic.

A call of *p-th* priority can be accommodated by the system only if the condition $C_{m,a} \geq C_{m,p}$ (for all the traffic classes of *m*=1… *M*) is satisfied after acceptance of a call of *p-th* priority traffic. Bandwidth is released from all of the existing bandwidth adaptive calls (*m*=1, 2, … , *M* ) to accommodate an arrival of a *p-th* priority traffic call. If $N_m$ is the number of existing calls of *m-th* class traffic, then overall releasable bandwidth from the system to accept a call of *p-th* priority traffic is

$$C_{releasable,p} = \sum_{m=1}^{M} N_m (C_{m,a} - C_{m,p}) \quad (3)$$

The maximum possible available bandwidth to accept a call of *p-th* priority traffic call is

$$C_{available,p} = C - \sum_{m=1}^{M} N_m C_{m,p} \quad (4)$$

From the definition it can be stated that,

$$1 > \gamma_{m,0} \geq \gamma_{m,1} \geq \cdots \geq \gamma_{m,m} \geq \cdots \geq \gamma_{m,M} \geq 0 \quad (5)$$

From (1) – (5) it is clearly observed that,

$$C_{m,0} \leq C_{m,1} \leq \cdots \leq C_{m,m} \leq \cdots \leq C_{m,M} \quad (6)$$

Equations (1) – (6) indicate that more bandwidth is releasable and available to accept a higher priority call. Therefore, by re-allocating the bandwidth of the bandwidth adaptive traffic, the proposed multi-level bandwidth allocation scheme can accept more calls of higher priority without reducing the bandwidth utilization. A call (of any class of traffic) can be accepted only if the required bandwidth for that call is less than or equal to the unused bandwidth plus releasable bandwidth.

Whenever the requested bandwidth is less than or equal to the total available bandwidth, the system accepts the call. Otherwise, the system calculates the minimum required bandwidth, maximum releasable bandwidth, and the maximum possible available bandwidth to accept the requesting call. Fig. 2 explains the system conditions to accept a call request in terms of present value of bandwidth degradation factor ($\gamma_m$). Based on the present value of bandwidth degradation factor, $\gamma_m$, the system can decide which types of requested calls can be accepted. The increased number of system calls increases the value of $\gamma_m$ for the adaptive multimedia traffic. This results in an increase of the blocking rate of lower priority traffic calls.

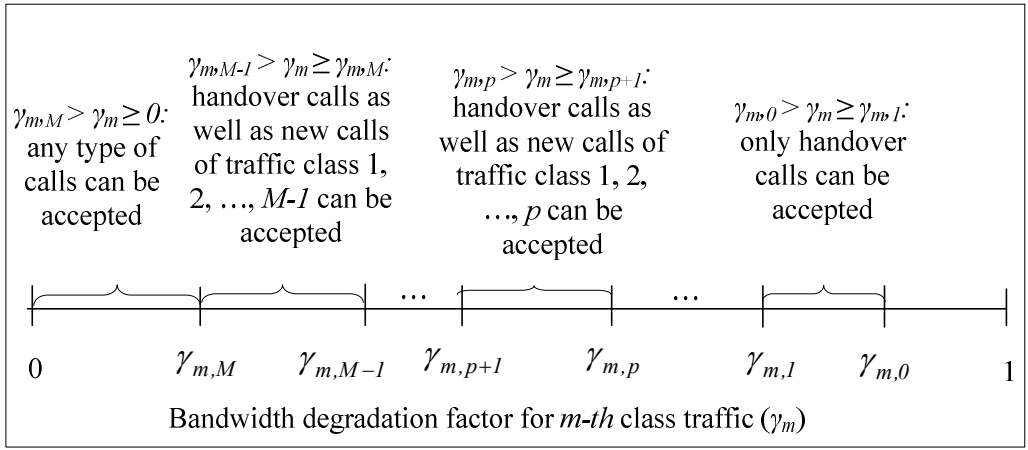

**Fig. 2** Description of system conditions to accept a call request in terms of present value of bandwidth degradation factor ($\gamma_m$).

Our proposed scheme is modeled as an *M/M/N/N* queuing system. The Markov Chain for the queuing analysis of the proposed scheme is shown in Fig. 3. We define $1/\mu$ as the average channel holding time (exponentially distributed). For the states between 0 and *N*, the bandwidths for existing calls are not degraded (hard-QoS). When the system traffic is increased, the system allows the bandwidth adaptation to admit more calls in the system. However, the number of permitted increased states for a traffic class is based on the priority of the traffic calls. In this figure, $\lambda_p$ represents the call arrival rate of *p-th* priority traffic calls. $\lambda_T$ denotes total call arrival rate. The maximum number of states of the Markov Chain in which the system accepts calls of *p-th* priority traffic is $N_p$. For the system states between 0 and *N* i.e., hard-QoS states, when there is enough bandwidth in the system, all the *M* traffic classes calls are allocated with the requested bandwidth. Thus, in these states, the average call duration time is same, $1/\mu$. However, when the state of the proposed system is more than $N$ i.e., during the soft-QoS states, bandwidth allocations for few or all of the existing calls are already degraded. Since the average call duration for some of the bandwidth adaptive calls (e.g., file transfer) depends on the bandwidth allocations for them, the average call duration time $1/\mu_i$ now depends on the state that the system $i$. Therefore, for the states above $N$, the average call duration time $1/\mu$ is greater than that of $1/\mu$ during the states between 0 and $N$.

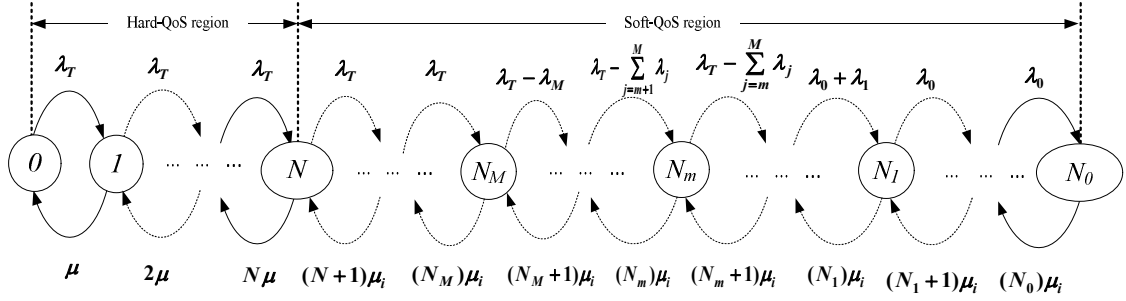

**Fig. 3** The Markov Chain for the proposed scheme.

The probability that the system is in state $i$, is given by $P_i$. From the Fig. 3, the state balance equations are expressed as

$$\begin{cases} i\mu P_i = \lambda_T P_{i-1}, & 0 \leq i \leq N \\ i\mu_i P_i = \lambda_T P_{i-1}, & N < i \leq N_M \\ i\mu_i P_i = (\lambda_T - \lambda_M) P_{i-1}, & N_M < i \leq N_{M-1} \\ i\mu_i P_i = (\lambda_T - \sum_{j=m}^{M} \lambda_j) P_{i-1}, & N_m < i \leq N_{m-1} \\ i\mu_i P_i = (\lambda_0 + \lambda_1) P_{i-1}, & N_2 < i \leq N_1 \\ i\mu_i P_i = \lambda_0 P_{i-1}, & N_1 < i \leq N_0 \end{cases} \quad (7)$$

A call of *m-th* class traffic is blocked in the proposed scheme if the state of the system calls is $N_m$ or larger. $N_M$ is the maximum number of available states for the calls of lowest priority (*M-th* class) traffic class. The highest priority is given to handover calls. The generalized equations to calculate the new call blocking probability for any call of traffic class $m$ among total $M$ number of traffic classes are derived using the queuing analysis. The handover call dropping probability ($P_D$) is calculated using (8). The new call blocking probability for the traffic class of $m$ is calculated using (9). In (9), $P_{B,0}=P_D$.

$$P_D = P_{N_0} = P_0 \frac{\lambda_T^{N_M}}{(N_0)! \mu^N \prod_{i=N+1}^{N_0} \mu_i} \prod_{k=1}^{M} (\lambda_0 + \lambda_1 + \lambda_2 + \ldots + \lambda_{M-k})^{N_{M-k}-N_{M-k+1}} \quad (8)$$

$$P_{B,m} = \sum_{i=N_m}^{N_0} P_i$$
$$= P_{B,m-1} + P_0 \sum_{i=N_m}^{N_{m-1}-1} \frac{\lambda_T^{N_M}}{i!(\mu_i)^i} (\lambda_0 + \lambda_1 + \lambda_2 + \ldots + \lambda_{m-1})^{i-N_m} \prod_{k=m}^{M} (\lambda_0 + \lambda_1 + \lambda_2 + \ldots + \lambda_k)^{N_k-N_{k+1}} \quad (9)$$

$$P_0 = \left[ 1 + \sum_{i=1}^{N_M} \frac{\lambda_T^i}{\mu^i i!} + \sum_{j=1}^{M} \sum_{i=N_j+1}^{N_{j-1}} \left\{ \frac{\lambda_T^{N_M}(\lambda_0 + \lambda_1 + \lambda_2 + \ldots + \lambda_{j-1})^{i-N_j}}{i!(\mu_i)^i} \prod_{k=j}^{M} (\lambda_0 + \lambda_1 + \lambda_2 + \ldots + \lambda_k)^{N_k-N_{k+1}} \right\} \right]^{-1} \quad (10)$$

# 3. Performance Analysis

In this section, we present the numerical results of the analysis of the proposed scheme. We compared the performance of our proposed scheme with several other schemes. Table 1 shows the basic assumptions for the traffic classes. We consider four classes of traffic calls. Handover calls are given higher priority over the new calls of any class of traffic. The requested call sequence is assumed to be random. The ratio of the number of requested new calls (voice: web-browsing: video: background) is considered as 3:3:1:2. From the calculation [10], by using user's average speed of 7.5 km/hr, cell radius of 1 km, and average call duration of 120 sec during the hard-QoS states, the average cell dwell time is found to be 240 sec. During the soft-QoS states, we consider state dependent average call duration i.e., more than 120 sec. Poisson distribution is assumed for the call arriving process. We assume that the system capacity is 6 Mbps.

**Table 1** Basic assumptions for the analysis

| Traffic class (m) | Requested bandwidth by each call | $p$ | $\gamma_{m,0}$ | $\gamma_{m,1}$ | $\gamma_{m,2}$ | $\gamma_{m,3}$ | $\gamma_{m,4}$ |
|---|---|---|---|---|---|---|---|
| Conversational voice (m=1) | 32 kbps | 1 | 0 | 0 | 0 | 0 | 0 |
| Interactive web-browsing (m=2) | 120 kbps | 2 | 0.6 | 0.55 | 0.5 | 0.45 | 0.4 |
| Streaming video (m=3) | 256 kbps | 3 | 0.7 | 0.65 | 0.6 | 0.55 | 0.5 |
| Background (m=4) | 60 kbps | 4 | 0.8 | 0.75 | 0.7 | 0.65 | 0.6 |

The allocated bandwidth for each of the bandwidth adaptive traffic calls is reduced with the increase of traffic congestion to admit more number of calls in the system. Fig. 4 shows the change of bandwidth allocation for each of the calls with the increase of traffic congestion in our proposed scheme. The result shows that the system provides requested bandwidth for all of the calls until the call arrival rate of 0.4 calls/sec. Therefore, the traffic load until this arrival rate can be treated as the light load condition and congestion occurs after this arrival rate (i.e., heavy load condition). The allocated bandwidth for each of the bandwidth adaptive calls is decreased with the increase of traffic load just after the call arrival rate of 0.4 calls/sec to admit more calls. The conversational voice calls are assumed to be bandwidth non-adaptive. Therefore, the system always provides constant bandwidth for each of the existing voice calls. Fig. 5 shows a comparison of the total releasable amount of bandwidth from the existing bandwidth adaptive calls to accept different types of call requests in our proposed scheme. At light load condition, the number of existing calls is small. Therefore, total releasable amount of bandwidth from the existing calls is increased with the increase of traffic load during light load condition. However, total releasable amount of bandwidth from the existing calls is decreased with the increase of traffic load during the soft-QoS condition (i.e., after the call arrival rate of 0.4 calls/sec).

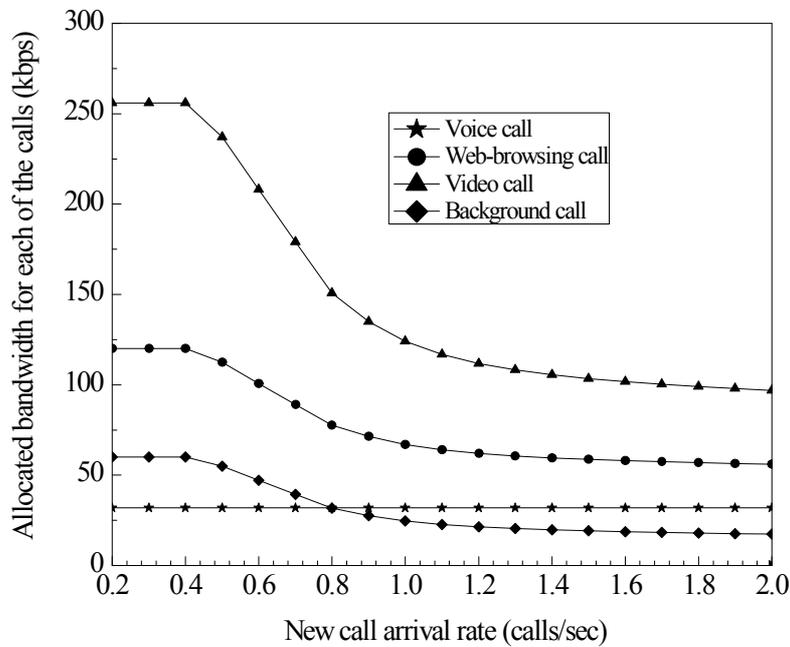

**Fig. 4** Bandwidth allocation for each of the existing calls.

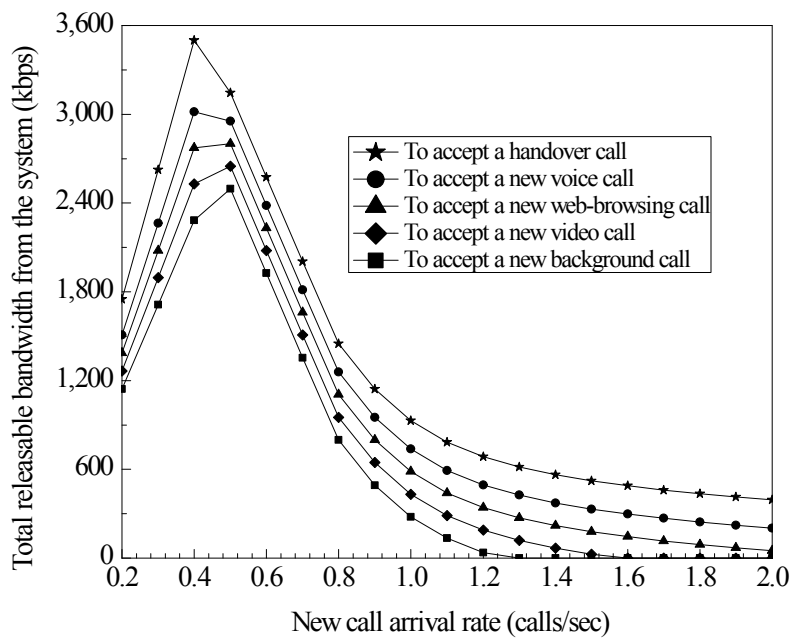

**Fig. 5** A comparison of total releasable amount of bandwidth from the existing bandwidth adaptive calls.

Fig. 6 shows that the proposed multi-level bandwidth adaptation scheme provides negligible handover call dropping probability even for highly congested traffic condition. It shows that our scheme offers a handover call dropping probability which is even less than 0.0002 at the call arrival rate of 2 calls/sec (i.e., during highly congested traffic condition). The proposed scheme also gives priority for the higher priority new calls over lower priority new calls. Therefore, the proposed scheme blocks few more new calls of lower priority if the bandwidth adaption is not sufficient to reduce the call blocking probability of the higher priority traffic calls. Fig. 7 shows that the proposed scheme provides significantly reduced new call blocking probabilities for the higher priority traffic calls during the congested traffic condition. The figure clearly demonstrate that at the call arrival rate of 1 call/sec, the proposed scheme offers 500 times less new call probability for the high priority voice calls compared to the new call probability of low priority background calls (i.e., 0.0005 for voice calls and 0.25 for background calls). The provision of maximum level of bandwidth adaptation without the priority of calls cannot offer satisfactory level of handover call dropping probability and new call blocking probability of the higher priority traffic calls. The non-adaptive non-priority scheme creates very high handover call dropping probability and new call blocking probability of the higher priority traffic calls. However, proper choice of the bandwidth degradation factors in the proposed scheme can avoid the increase of overall forced call termination probability compared to the non-priority bandwidth adaptive scheme. Fig. 8 shows that our proposed scheme provides almost equal forced call termination probability compared to the adaptive non-priority scheme. Therefore, the priorities of handover calls and other calls do not increase the overall forced call termination probability significantly. Whereas the non-adaptive non-priority scheme results in very high overall forced call termination probability.

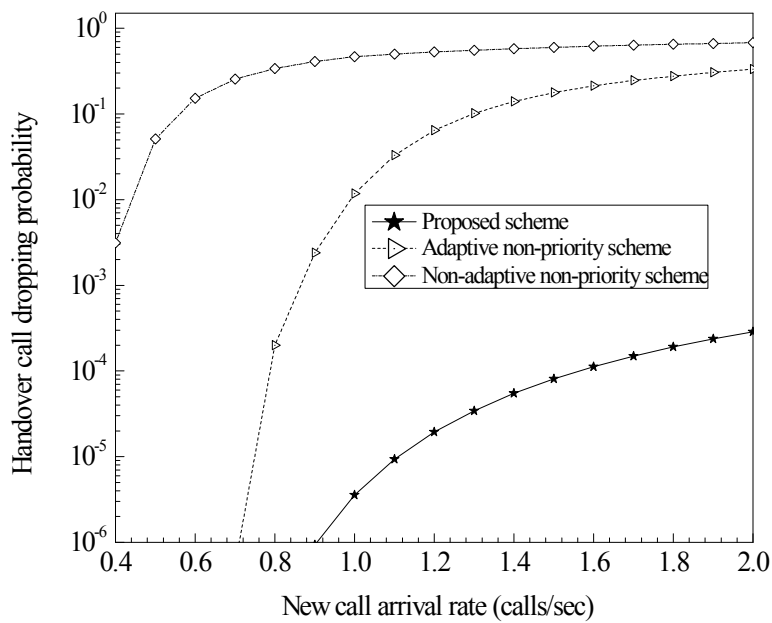

**Fig. 6** A comparison of handover call dropping probability.

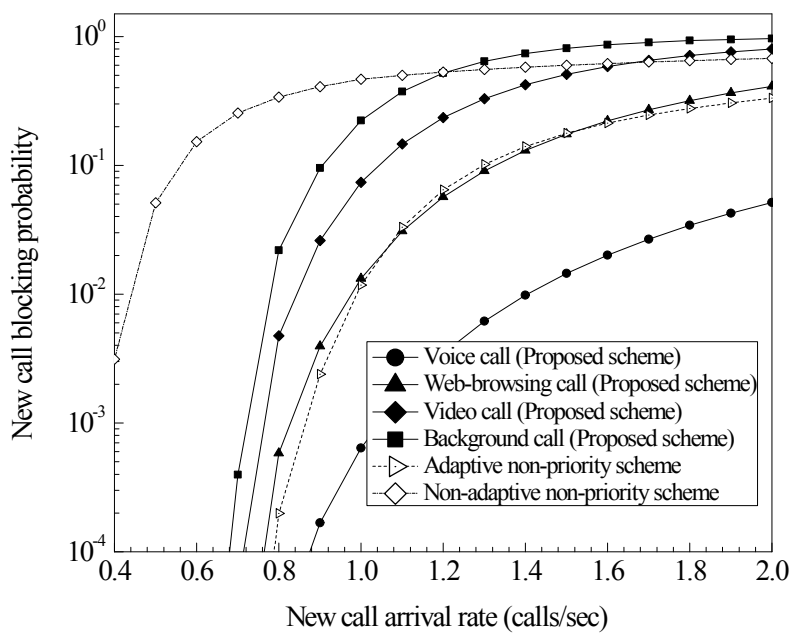

**Fig. 7** A comparison of new call blocking probability.

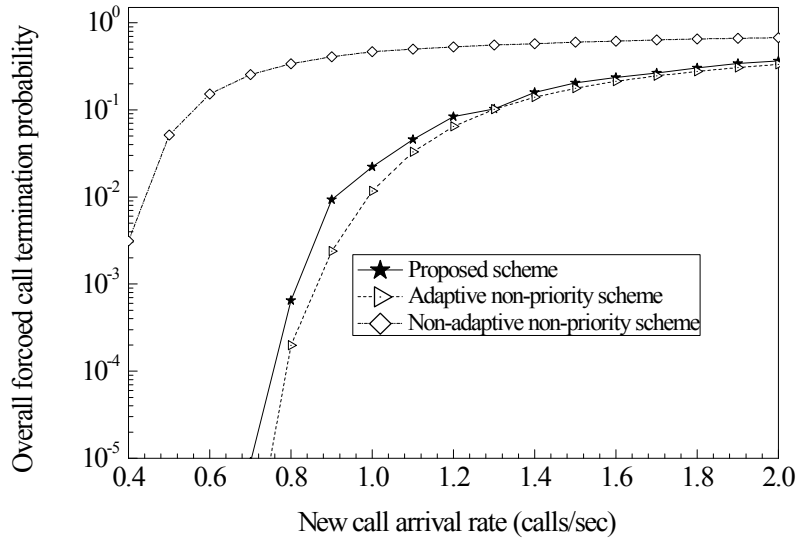

Fig. 8. A comparison of overall forced call termination probability.

Figs. 6-8 clearly show the advantages of the proposed scheme. The proposed scheme can maximize the number of call admission as well as can guarantee the service connectivity for the higher priority traffic calls.

## 4. Conclusions

In this paper, we proposed a multi-level bandwidth adaptation for priority based call admission in wireless networks. The scheme releases multi-level of bandwidth from already admitted bandwidth adaptive calls as to accommodate handover calls and multi-class new calls when available bandwidth is low. The proposed scheme offers more bandwidth degradation of the existing and requesting calls to support higher priority traffic calls over lower priority calls. Therefore, the proposed scheme provides priority of traffic calls to offer better service connectivity for the higher priority users as well as bandwidth adaptation to maximize the number of call admission. As a result, to give the priority of traffic calls, the overall forced call termination probability is not increased significantly. While employing the proposed scheme, the network operator has the opportunity to control the minimum QoS level for each of the traffic classes and the desired level of handover call dropping probability as well as the new call blocking probabilities.

The results based on a novel theoretical model presented in this paper were studied using several numerical analyses. Experimental results for comparison to theory are saved for future research work.

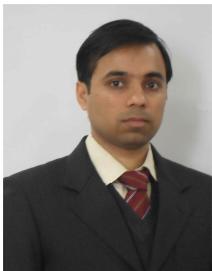

**Mostafa Zaman Chowdhury** received his B.Sc. degree in electrical and electronic engineering from Khulna University of Engineering and Technology (KUET), Bangladesh, in 2002. He received his M.Sc. and Ph.D. degrees both in electronics engineering from Kookmin University, Korea, in 2008 and 2012, respectively. In 2003, he joined the Electrical and Electronic Engineering Department at KUET as a faculty member. Currently he is working as an Assistant Professor at the same department. In 2008, he received the Excellent Student Award from Kookmin University. One of his papers received the Best Paper Award at the International Conference on Information Science and Technology in April 2012 in Shanghai, China. He served as a reviewer for several international journals (including IEEE Communications Magazine, IEEE Transaction on Vehicular Technology, IEEE Communications Letters, IEEE Journal on Selected Areas in Communications, Wireless Personal Communications (Springer), Wireless Networks (Springer), Mobile Networks and Applications (Springer), and Recent Patents on Computer Science) and IEEE conferences. He has been involved in several Korean government projects. His research interests include convergence networks, QoS provisioning, mobility management, femtocell networks, and VLC networks.

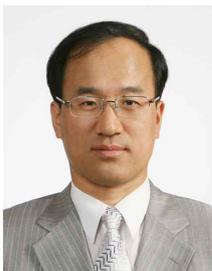

**Yeong Min Jang** received the B.E. and M.E. degrees both in electronics engineering from Kyungpook National University, Korea, in 1985 and 1987, respectively. He received the doctoral degree in Computer Science from the University of Massachusetts, USA, in 1999. He worked for ETRI between 1987 and 2000. Since September 2002, he is with the School of Electrical Engineering, Kookmin University, Seoul, Korea. He has organized several conferences such as ICUFN2009, ICUFN2010, ICUFN2011, ICUFN2012, and ICUFN2013. He is currently a member of the IEEE and a life member of KICS (Korean Institute of Communications and Information Sciences). He had been the director of the Ubiquitous IT Convergence Research Center at Kookmin University since 2005 and the director of LED Convergence Research Center at Kookmin University since 2010. He has served as the executive director of KICS since 2006. He had been the organizing chair of Multi Screen Service Forum of Korea since 2011. He had been the Chair of IEEE 802.15 LED Interest Group (IG-LED). He received the Young Science Award from the Korean Government (2003 to 2005). He had served as the founding chair of the KICS Technical Committee on Communication Networks in 2007 and 2008. His research interests include 5G mobile communications, radio resource management, small cell networks.